\documentclass[twocolumn,showpacs,aps,pra]{revtex4}

\newcommand{\ii}{\mathrm i}

\usepackage{epsf}
\usepackage{amsmath}
\usepackage{amsfonts}
\usepackage{amssymb}
\usepackage{bm}
\usepackage{bbm}
\usepackage{nicefrac}
\usepackage{color}
\usepackage{pifont}
\usepackage{graphicx}
\usepackage{enumerate}
\usepackage{dcolumn}
\usepackage{maybemath}

\begin{document}

\title{Fine--Structure Constant for Gravitational and Scalar Interactions}

\author{U.~D.~Jentschura}

\affiliation{Department of Physics, Missouri University of Science and Technology,
Rolla, Missouri, MO65409-0640, USA}

\affiliation{MTA--DE Particle Physics Research Group,
P.O.Box 51, H--4001 Debrecen, Hungary}

\begin{abstract}
Starting from the coupling of a relativistic quantum particle to the curved
Schwarzschild space-time, we show that the Dirac--Schwarzschild problem has
bound states and calculate their energies including relativistic
corrections. Relativistic effects are shown to be suppressed by the
gravitational fine-structure constant $\alpha_G = G \, m_1 \, m_2/(\hbar c)$,
where $G$ is Newton's gravitational constant, $c$ is the speed of light and
$m_1$ and $m_2 \gg m_1$ are the masses of the two particles. The kinetic corrections
due to space-time curvature are shown to lift the familiar $(n,j)$
degeneracy of the energy levels of the hydrogen atom.
We supplement the discussion by a consideration of an
attractive scalar potential, which,
in the fully relativistic Dirac formalism,
modifies the mass of the particle according to the replacement
$m \to m ( 1 - \lambda/r )$, where $r$ is the radial coordinate.
We conclude with a few comments regarding the $(n,j)$ degeneracy 
of the energy levels, where $n$ is the principal quantum number, 
and $j$ is the total angular momentum, and illustrate the calculations
by way of a numerical example.
\end{abstract}


\maketitle

%
%
\section{Introduction}
\label{sec1}

As one combines relativistic quantum mechanics~\cite{Di1928a,Di1928b} with
general relativity~\cite{Ei1915,Hi1915,Ei1916}, one has to formulate the Dirac
equation on a curved
space-time~\cite{BrWh1957,Bo1975prd,SoMuGr1977,SiTe2005,Je2013,JeNo2013pra,JeNo2014jpa}.
One of the most paradigmatic calculations concerns the Dirac--Schwarzschild
Hamiltonian~\cite{Sc1916,Ed1924,SiTe2005}, which is obtained for a Dirac
particle in the static Schwarzschild metric. The Dirac--Schwarzschild problem
constitutes the analogue of the Dirac--Coulomb
problem~\cite{SwDr1991a,SwDr1991b,SwDr1991c}, which is otherwise relevant for
the Dirac particle bound to a central Coulomb potential, as opposed to a
central gravitational field.  The main problem is that, unlike for the
Dirac--Coulomb problem, the gravitational central-field Dirac--Schwarzschild
problem cannot be treated based on the correspondence principle alone. 

Namely, the gravitational potential $-G \, m_1\, m_2/r$ cannot simply be
inserted into the Dirac--Schwarzschild Hamiltonian. One first has to
couple~\cite{BrWh1957,Bo1975prd,SoMuGr1977} the Dirac particle to the curved
space-time, using a fully covariant formalism, and then, identify the
translation operator for the time coordinate with the Dirac Hamiltonian.  This
identification becomes unique in the Dirac--Schwarzschild problem when we
demand that the time coordinate have a smooth limit to the flat-space time in
the regime of large separation~\cite{Ed1924,Je2013,JeNo2013pra,JeNo2014jpa}.

We recall that for the Dirac--Coulomb problem, one simply adds the Coulomb
potential $-Z e^2/(4 \pi \epsilon_0 \, r)$ to the free Dirac Hamiltonian, in
the sense of a minimal coupling of the bound electron to the central
electrostatic field of the nucleus~\cite{SwDr1991a,SwDr1991b,SwDr1991c}. Here,
$Z$ is the nuclear charge number, $e$ is the elementary charge, $\epsilon_0$ is
the vacuum permittivity, and $r$ is the distance from the center of the
potential. Both the Dirac--Schwarzschild as well as the Dirac--Coulomb
Hamiltonians take into account the gauge boson exchange (graviton exchange and
Coulomb photon exchange, respectively) to all orders, but only in the classical
approximation.  This is sufficient to calculate the corrections of order
$\alpha_G^4$ and $\alpha_{\rm QED}^4$, where $\alpha_G^4$ and $\alpha_{\rm
QED}^4$ denote the gravitational and electrodynamic fine-structure constants,
respectively.

We anticipate that the familiar $(n,j)$ degeneracy of the energy levels of the
Dirac--Coulomb problem will be lifted for gravitational coupling, which 
implies that for example, the
gravitationally coupled $2S$ and $2P_{1/2}$ levels are not degenerate. 
For the electromagnetically coupled hydrogen atom,
the corresponding degeneracy is lifted only by the Lamb shift;
the theoretical explanation involves a manifestly 
quantized electromagnetic field~\cite{Be1947}. The reason for
the lifted degeneracy, in the case of gravitational coupled,
is different: Namely,we observe that it is due to the space-time curvature
corrections to the kinetic term in the Dirac--Schwarzschild Hamiltonian.  This
finding is illustrated by a comparison to the energy levels of an attractive
scalar potential, which are also calculated here, including relativistic
corrections.

This paper is organized as follows. In Sec.~\ref{sec2}, we consider the
fine structure of the energy levels of the Dirac--Schwarzschild Hamiltonian and
express the result in terms of the gravitational fine-structure constant
$\alpha_G$, and of the quantum numbers of the bound state.  In passing, we
clarify that the quantum mechanical gravitational central-field problem has
bound states. For clarity, but without loss of generality, we consider a
gravitationally coupled ``atom'' consisting of electron and proton. In
Sec.~\ref{sec3}, we compare to the energy levels of an attractive scalar
potential.  Having clarified the physical origin of the correction terms which
lift the $(n,j)$ degeneracy, we continue in Sec.~\ref{sec4} with the
identification of a set of physical parameters for a gravitationally coupled
system, where the calculations reported here might be phenomenologically
relevant. These concern an electron gravitationally coupled,
in a Rydberg state, to a black hole of mass $10^{-11} \, M_E$,
where $M_E$ is the mass of the Earth.
In the derivations, we use the electron mass
$m_e$ and the proton mass $m_p$, Newton's gravitational constant $G$, Planck's
reduced quantum unit of action $\hbar$, and the speed of light $c$.  Units with
$\hbar = c = \epsilon_0 = 1$ are used in this paper unless explicitly stated
otherwise (in some manipulations, it will be of advantage to temporarily switch
back to the SI mksA unit system).

%
%
\section{Dirac--Schwarzschild Fine Structure}
\label{sec2}

We start from the Dirac--Schwarzschild Hamiltonian $H$
for particle of mass $m_e$ in the central gravitational
field of a particle (or planet) of mass $m_p \gg m_e$
(see Ref.~\cite{JeNo2013pra}),
\begin{equation}
H = \frac12 \, \left\{ \vec\alpha \cdot \vec p,
\left( 1 - \frac{G \, m_p}{2 r} \right) \right\} +
\beta \, m_e \left( 1 - \frac{G \, m_p}{r} \right) \,,
\end{equation}
The mass parameters $m_e$ and $m_p$ are canonically associated with the
electron and proton masses.  However, the considerations reported in the
following remain valid, without loss of generality, for any small mass $m_1 =
m_e$ in the gravitational field of a larger, central mass $m_2 = m_p$.  The
nonrecoil approximation is employed.  The 
vector of the Dirac $\vec\alpha$ matrices and the 
Dirac  $\beta$ matrix  are used in the standard
representation~\cite{SwDr1991a,SwDr1991b,SwDr1991c,Je2013,JeNo2013pra}.

After a Foldy--Wouthuysen transformation~\cite{FoWu1950},
one obtains the Dirac--Schwarzschild Hamiltonian $H_{\rm DS}$.
It is characterized by an overall 
prefactor matrix $\beta$, which expresses the particle--antiparticle 
symmetry inherent to the gravitationally 
coupled Dirac theory [see Eq.~(28) of Ref.~\cite{SiTe2005}
and Eq.~(21) of Ref.~\cite{JeNo2013pra} for a manifestly 
Hermitian form].
In order to obtain the leading 
relativistic corrections, one may 
restrict the wave function to the 
``upper'' two-component spinor,
and the Dirac--Schwarzschild Hamiltonian $H_{\rm DS}$
to its upper $(2 \times 2)$-submatrix,
%
\begin{align}
\label{HDS}
& H_{\rm DS} = 
\frac{\vec p^{\,2}}{2 m_e} - \frac{G m_e m_p}{r}
- \frac{\vec p^{\,4}}{8 m_e^3}
\\[0.133ex]
& \; - \frac{3 G m_p}{4 m_e} \left\{ \vec p^{\,2}, \frac{1}{r} \right\}
+ \frac{3 \pi G m_p }{2 m_e} \, \delta^{(3)}(\vec r)
+ \frac{3 G m_p \, \vec\sigma \cdot \vec L}{4 m_e r^3} \,.
\nonumber
\end{align}
The vector of $(2 \times 2)$--Pauli matrices 
is denoted as $\vec\sigma$.
The momentum operator in Eq.~\eqref{HDS} 
is given as $\vec p = -\ii \hbar \vec\nabla_r$,
where we temporarily restore SI mkSA units
for absolute clarity.
We employ the following scaling to dimensionless
quantities $\rho$,
\begin{subequations}
\begin{align}
\label{rrho}
r =& \; \frac{\hbar^2}{G \; m_e^2 \; m_p} \, \rho \,,
\qquad
\vec\nabla_r = \frac{G \; m_e^2 \; m_p}{\hbar^2} \, \vec\nabla_\rho \,,
\\[0.133ex]
\vec p =& \; -\ii \, \frac{G \; m_e^2 \; m_p}{\hbar} \, \vec\nabla_\rho \,.
\end{align}
\end{subequations}
Here, $\vec\nabla_\rho$ is the dimensionless gradient operator,
with respect to the dimensionless coordinate $\rho$. 
The scaled leading-order term has the Schr\"{o}dinger-like structure
\begin{equation}
H_S = \frac{\vec p^{\,2}}{2 m_e} - \frac{G m_e m_p}{r} 
= \alpha_G^2 m_e  c^2 \, \left( -\frac12 \, \vec\nabla^2_\rho - 
\frac{1}{\rho} \right) \,.
\end{equation}
For the electron-proton system, 
employing the {\tt CODATA}~\cite{MoTaNe2012}  value of 
$G = 6.67384(80) \times 10^{-11} \, {\rm N} \,
\frac{{\rm m}^2}{{\rm kg}^2}  $, one obtains
\begin{equation}
\label{alphaG}
\alpha_G = \frac{G \, m_e \, m_p}{\hbar \, c} = 
3.21637(39) \times 10^{-42} \,.
\end{equation}
Today, Newton's gravitational constant $G$
remains~\cite{MoTaNe2012} one of the least well known physical constants to date,
with a relative uncertainty of $1.2 \times 10^{-4}$.
We should note that the numerically small value of the 
gravitational fine-structure constant $\alpha_G$ given in Eq.~\eqref{alphaG}
is tied to the physical system under consideration,
namely, the electron-proton system. 
The gravitational Bohr radius of the electron-proton 
system is 
\begin{equation}
a_{0,G} = \frac{\hbar^2}{G \; m_e^2 \; m_p} \approx 
1.20 \times 10^{29} \, {\rm m} \,,
\end{equation}
which is very large but depends on the masses employed.
For other systems composed of elementary particles 
or black holes of various masses,
the value of the gravitational fine-structure constant 
is different. One may remark that Eddington~\cite{Ed1931} observes
that the electromagnetic fine-structure constant $\alpha_{\rm QED} 
\approx 1/137.036$ and the gravitational fine-structure 
constant $\alpha_G^{(ee)}$ for two gravitationally 
interaction electrons fulfill the approximate numerical
relationship
\begin{equation}
\label{eddington}
\frac{\alpha_{\rm QED}}{\alpha_G^{(ee)}} =
\frac{e^2}{4 \pi \epsilon_0 G \, m^2_e} \approx 4.2 \times 10^{42}  
\approx \sqrt{ \, N_C \,} \,,
\end{equation}
where $N_C$ is the number of charged particles in the Universe.
We shall not comment on this numerical coincidence here
except for reemphasizing that the gravitational interactions
of elementary particles are much weaker than 
electromagnetic and ``weak'' interactions, as well as
strong interactions. Still, to fix ideas, it is instructive
to consider that bound electron-proton system, 
The Schr\"{o}dinger eigenenergies 
of the eigenproblem $H_S | \phi \rangle = E_n | \phi \rangle$ 
are given as follows,
\begin{equation}
E_n = -\frac{\alpha_G^2 m_e c^2}{2 n^2} \,.
\end{equation}
For the relativistic correction term given in Eq.~\eqref{HDS},
it is instructive to consider the 
scaling of the various relativistic 
correction terms separately,
with full reference to the SI mksA unit system,
\begin{subequations}
\begin{align}
& \; - \frac{\vec p^{\,4}}{8 \, m_e^3 \, c^2} =
- \frac{\hbar^4 \, \vec\nabla^4_r}{8 \, m_e^3 \, c^2}  =
- \frac18 \, \alpha_G^4 \, m_e \, c^2 \, \vec\nabla^4_\rho \,,
\\[0.133ex]
& \; - \frac{\hbar^2 }{c^2} \, \frac{3 G m_p }{4 m_e} \, 
\left\{ \vec p^{\,2}, \frac{1}{r} \right\} =
\frac34 \, \alpha_G^4 \, m_e \, c^2 \,
\left\{ \vec\nabla^2_\rho, \frac{1}{\rho} \right\} \,,
\\[0.133ex]
& \; 
\frac{\hbar^2 }{c^2} \,
\frac{3 \pi G m_p }{2 m_e} \, \delta^{(3)}(\vec r) =
\alpha_G^4 \, m_e \, c^2 \,
\frac{3 \, \pi}{2} \, \delta^{(3)}(\vec \rho) \,,
\\[0.133ex]
& \; \frac{\hbar^2 }{c^2} \,
\frac{3 G m_p \, \vec\Sigma \cdot \vec L}{4 m_e \, r^3} =
\alpha_G^4 \, m_e \, c^2 \, 
\frac{3 \, \vec\Sigma \cdot \vec L}{4 \, \rho^3} \,.
\end{align}
\end{subequations}
These considerations manifestly identify the relativistic 
correction terms to be of order $\alpha_G^4$.
The scaled Dirac--Schwarzschild Hamiltonian 
with relativistic corrections thus is given as follows,
\begin{align}
\label{HDSscaled}
& H_{\rm DS} = \alpha_G^2 m_e  c^2 \, \left( -\frac12 \, \vec\nabla^2_\rho - 
\frac{1}{\rho} \right) + \alpha_G^4 \, m_e \, c^2
\\[0.133ex]
& \; 
\times \left( 
- \frac18 \, \vec\nabla^4_\rho 
+ \frac34 \, \left\{ \vec\nabla^2_\rho, \frac{1}{\rho} \right\} 
+ \frac{3 \, \pi}{2} \, \delta^{(3)}(\vec \rho) 
+ \frac{3 \, \vec\Sigma \cdot \vec L}{4 \, \rho^3} \right) \,.
\nonumber
\end{align}
Using formulas given on p.~17 of Ref.~\cite{BeSa1957},
we may evaluate the relativistic corrections 
as a function of the bound-state quantum numbers
($n$ is the principal quantum number, $\ell$ is the 
orbital angular momentum quantum number, and 
$j$ is the total angular momentum quantum number).
The calculation proceeds via first-order 
perturbation theory, starting from the Schr\"{o}dinger--Pauli 
wave function $\psi_{n \ell j}(\vec \rho) = R_{n\ell}(\rho) \,
\chi_{\varkappa \mu}(\hat \rho)$, where 
\begin{equation}
\label{varkappa}
\varkappa = 2 (\ell - j) \, (j + 1/2)
= (-1)^{j+l+1/2} \, \left(j + \frac12 \right)
\end{equation}
is the Dirac angular quantum number~\cite{Ro1961,SwDr1991a}.
Some exemplary radial parts $R_{n\ell}(\rho)$ of the 
Schr\"{o}dinger--Pauli wave functions are given in 
p.~15 of Ref.~\cite{BeSa1957}.
Knowing $j$ and $\ell$, one may calculate 
$\varkappa$ using Eq.~\eqref{varkappa}.
Conversely, one may calculate $\ell$ 
with the help of the formula
$\ell = | \varkappa + 1/2 | - 1/2$.
The relativistic corrections amount to
\begin{align}
\label{mainres}
E_{n \ell j} =& \; -\frac{\alpha_G^2 m_e c^2}{2 n^2} +
\alpha^4_G m_e c^2 \, 
\left( \frac{15}{8 n^4} \right.
\nonumber\\[0.133ex]
& \; - \frac{7 j + 5}{(j+1) \, (2j+1) \, n^3} \, \delta_{\ell, j+1/2}
\nonumber\\[0.133ex]
& \; \left. 
- \frac{7 j + 2}{j \, (2j+1) \, n^3} \, \delta_{\ell, j-1/2} \right) \,.
\end{align}
The $S$ state energy can be obtained from Eq.~\eqref{mainres}
with the help of the term $\ell = 0$ and $j = 1/2$;
$S$ states are the only ones for which the 
expectation value of the Dirac-$\delta$ term in 
Eq.~\eqref{HDSscaled} is nonvanishing;
the result reads as
\begin{align}
E_{n S_{1/2}} =& \; 
-\frac{\alpha_G^2 m_e c^2}{2 n^2} +
\alpha^4_G m_e c^2 \,
\left( \frac{15}{8 n^4} - \frac{11}{2 n^3} \right) \,.
\end{align}
The $2S_{1/2}$, $2P_{1/2}$ and $2P_{3/2}$ levels are 
given as follows,
\begin{subequations}
\begin{align}
E_{2 S_{1/2}} =& \; -\frac18 \, \alpha_G^2 m_e c^2 - 
\frac{73}{128} \, \alpha^4_G m_e c^2 \,,
\\[0.133ex]
E_{2 P_{1/2}} =& \; -\frac18 \, \alpha_G^2 m_e c^2 - 
\frac{91}{384} \, \alpha^4_G m_e c^2 \,,
\\[0.133ex]
E_{2 P_{3/2}} =& \; -\frac18 \, \alpha_G^2 m_e c^2 - 
\frac{55}{384} \, \alpha^4_G m_e c^2 \,.
\end{align}
\end{subequations}
While there is no degeneracy, the hierarchy 
$ E_{2 S_{1/2}} < E_{2 P_{1/2}} < E_{2 P_{3/2}}$
follows a somewhat general paradigm of bound-state 
theory~\cite{BeJa1986}, namely, that 
states with higher angular momentum quantum numbers
have higher energy.

%
%
\section{Fine Structure for a Scalar Potential}
\label{sec3}

The Dirac Hamiltonian with a $(r/1)$-scalar potential~\cite{JeNo2014jpa} 
reads as follows (in natural units),
\begin{equation}
H = \vec \alpha \cdot \vec p + 
\beta \, \left( m - \frac{\lambda}{r} \right) \,,
\end{equation}
After the Foldy--Wouthuysen transformation, we have
\begin{align}
\label{HSP}
& H_{\rm SP}=
\beta \left( m+\frac{\vec p^{\,2}}{2m} 
- \frac{\lambda}{r} \right.
\\[0.133ex]
& \; - \left. \frac{\vec p^{\,4}}{8 m^3} 
+\frac{\lambda}{4m^2}\left\{\vec p^{\,2},\frac{1}{r}\right\}
-\frac{\pi \, \lambda}{2 m^2} \; \delta^{(3)}(\vec r) 
- \frac{\lambda \vec \Sigma \cdot \vec L }{4 m^2 r^3} \right) \,.
\nonumber
\end{align}
The scaling to dimensionless variables is analogous 
to Eq.~\eqref{rrho},
\begin{align}
\label{rrho_scalar}
r =& \; \frac{1}{\lambda \, m} \, \rho \,,
\qquad
\vec\nabla_r = m \, \lambda \, \vec\nabla_\rho \,,
\\[0.133ex]
\vec p =& \; -\ii \, m \, \lambda \, \vec\nabla_\rho \,,
\qquad
\alpha_S \equiv \lambda \,.
\end{align}
The role of the ``scalar fine-structure constant'' is 
taken by the variable $\alpha_S = \lambda$, 
and the scaled Hamiltonian reads as follows,
\begin{align}
& H_{\rm SP} =
\alpha_S^2 \, m \, \left( -\frac12 \, \vec\nabla^2_\rho - 
\frac{1}{\rho} \right) + \alpha_S^4 \, m
\\[0.133ex]
& \; \times 
\left( - \frac18 \, \vec\nabla^4_\rho 
- \frac14 \, \left\{ \vec\nabla^2_\rho, \frac{1}{\rho} \right\} 
- \frac{\pi}{2} \, \delta^{(3)}(\vec \rho) 
- \frac{\vec\Sigma \cdot \vec L}{4 \, \rho^3} \right) \,.
\nonumber
\end{align}
The energy levels are given as 
\begin{align}
\label{mainres_scalar}
E_{n \ell j} =& \; -\frac{\alpha_S^2 \, m \, c^2}{2 n^2} +
\alpha^4_S \, m \, c^2 \, 
\left( -\frac{1}{8 n^4} + \frac{1}{n^3 \, (j+1)} \right) \,.
\end{align}
Here, an important observation can be made:
In contrast to Eq.~\eqref{mainres},
the result for the relativistic corrections
of order $\alpha_S^4$ in the case of the scalar potential
has a compact functional form, and the $(n,j)$
degeneracy familiar from the Dirac--Coulomb 
problem (see Appendix~\ref{appa}) is restored. We also note that the 
Dirac--Schwarzschild Hamiltonian~\eqref{HDS} and 
the scalar Dirac Hamiltonian~\eqref{HSP} 
both entail ``$(1/r)$-modifications of the mass term'',
namely, the terms
\begin{equation}
\beta \, m_e \, \left( 1 - \frac{G \, m_p}{r} \right) 
\Leftrightarrow 
\beta \, m \, \left( 1 - \frac{\lambda}{r} \right) 
\end{equation}
However, in addition to this modification,
the Dirac--Schwarzschild Hamiltonian~\eqref{HDS} 
contains a modification of the kinetic term 
$\vec\alpha \cdot \vec p$ which is responsible for the 
lifting of the $(n,j)$ degeneracy,
as a comparison of Eqs.~\eqref{mainres} and~\eqref{mainres_scalar}
shows.

%
%
\section{Numerical Example}
\label{sec4}

\begin{figure}[t!]
\begin{center}
\begin{minipage}{1.0\linewidth}
\includegraphics[width=0.81\linewidth]{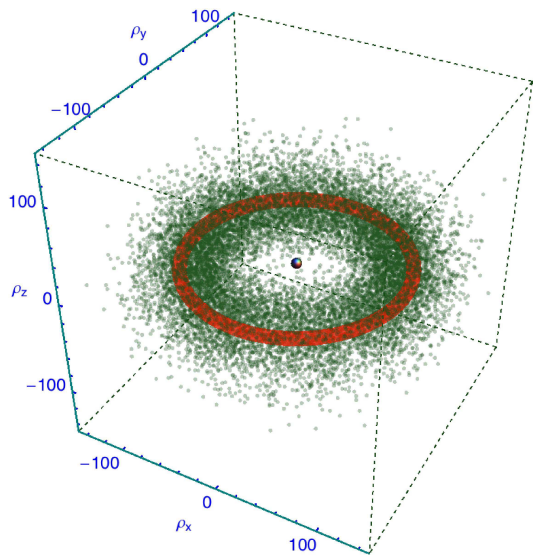}
\caption{\label{fig1} (Color.) Scatter plot 
of the probability density of an electron in a circular Rydberg state
with quantum numbers $n=10$, $\ell=m=9$, 
gravitationally bound to a black hole of mass $10^{-11}$ 
times the Earth mass. The green points are distributed 
according to the probability density $|\psi|^2$ of finding the 
electron at a particular point in space.
A corresponding circular classical 
trajectory with a radius of $105$
gravitational Bohr radii is indicated in red for comparison,
and the black hole at the center is indicated as a black dot.}
\end{minipage}
\end{center}
\end{figure}

Let us consider a ``tiny black hole'' of mass
$m_{\rm BH}$ to be $10^{-11}$ times the mass 
$M_E$ of the Earth, 
\begin{equation}
M_E \approx 5.9742 \times 10^{24} \, {\rm kg}\,,
\quad
m_{\rm BH} = 5.9742 \times 10^{13} \, {\rm kg}\,,
\end{equation}
and assume that the electric dipole polarizability of the very dense
black hole is vanishing.
The Schwarzschild radius $r_{S, {\rm BH}}$ is given as follows,
\begin{equation}
r_{S, {\rm BH}} = \frac{2 \, G \, m_{\rm BH}}{c^2} =
8.8731 \, \times 10^{-14} \, {\rm m} \,.
\end{equation}
The gravitational fine-structure constant
for an electron gravitationally bound to the 
black hole is given as 
\begin{equation}
\label{alphaGBH}
\alpha_{G,{\rm BH}} = 
\frac{G \, m_e \, m_{\rm BH}}{\hbar c} = 0.1148 \,.
\end{equation}
The gravitational Bohr radius is
\begin{equation}
a_{0,{\rm BH}} = \frac{\hbar^2}{G \, m_e^2 \, m_{\rm BH}} =
3.3612 \times 10^{-12} \, {\rm m} \,.
\end{equation}
In accordance with Eq.~\eqref{rrho}, we define the 
Cartesian components of the scaled dimensionless
coordinate $\vec \rho$ as follows,
\begin{equation}
\rho_x = \frac{x}{a_{0,{\rm BH}}} \,,
\qquad
\rho_y = \frac{x}{a_{0,{\rm BH}}} \,,
\qquad
\rho_z = \frac{x}{a_{0,{\rm BH}}} \,.
\end{equation}
In Fig.~\ref{fig1}, we present a ``scatter plot'' 
of the bound state with quantum numbers $n= 10$,
$\ell = 9$, and magnetic orbital angular momentum
projection $m = |\ell| = 9$ (``circular Rydberg state''),
where the points representing the wave function 
are distributed according to the 
probability density given by the absolute square of the 
wave function $|\psi|^2$.
The probability density of the Rydberg state inside the Schwarzschild radius 
 is  negligible and the expectation
value of the zitterbewegung term in the Dirac--Schwarzschild
Hamiltonian~\eqref{HDS} vanishes.
The nonrelativistic Schr\"{o}dinger--type approximation
is justified because the gravitational 
fine-structure constant $\alpha_{G,{\rm BH}}$ is small
against unity.
According to p.~17 of Ref.~\cite{BeSa1957},
the radial expectation value in the Schr\"{o}dinger state is 
$ \left< |\vec \rho| \right> = 105$ gravitational
Bohr radii. A classical circular trajectory circling the 
black hole is indicated in Fig.~\ref{fig1} for comparison.

According to Eq.~\eqref{mainres},
the bound-state energies for the two states
with $j = 9 \pm 1/2$ are given as follows,
\begin{subequations}
\begin{align}
\label{ea}
E_{n=10, \ell=9, j=19/2} =& \;
\left( - \frac{\alpha_{G,{\rm BH}}^2}{200} 
- \frac{263 \, \alpha_{G,{\rm BH}}^4}{1520000} \right) m_e c^2 \,,
\nonumber\\[0.133ex]
=& \; -33.7397 \, {\rm eV} \,,
\\[0.133ex]
\label{eb}
E_{n=10, \ell=9, j=17/2} =& \;
\left( - \frac{\alpha_{G,{\rm BH}}^2}{200} 
- \frac{173 \, \alpha_{G,{\rm BH}}^4}{912000} \right) m_e c^2 
\nonumber\\[0.133ex]
=& \; -33.7412 \, {\rm eV} \,,
\end{align}
\end{subequations}
The higher value of the total angular momentum 
$j$ moves the state with $j=19/2$ energetically upward.
Both energies~\eqref{ea} and~\eqref{eb} are numerically
close to the nonrelativistic approximation,
which reads as $-\alpha_{G,{\rm BH}}^2 m_e c^2/200= -33.7243 \, {\rm eV}$.

%
%
\section{Conclusions}
\label{sec5}

We have divided the current paper into three parts, the first of which (see
Sec.~\ref{sec2}) deals with the leading-order relativistic corrections to the
energies of bound states of the Dirac--Schwarzschild Hamiltonian~\eqref{HDS},
while the second part (Sec.~\ref{sec3}) investigates the bound states of a
Dirac Hamiltonian with a scalar $(1/r)$-potential. The latter potential
modifies the mass term of the Dirac particle; it is commonly referred to as a
scalar potential because of its properties under Lorentz
transformations~\cite{JeNo2014jpa}.  Having clarified the origin of the terms
that lift the $(n,j)$ degeneracy otherwise observed for scalar Dirac bound
states and for the Dirac--Coulomb problem (see Sec.~\ref{sec3} and
Appendix~\ref{appa}, respectively), we then turn our attention back to the
Dirac--Schwarzschild problem in Sec.~\ref{sec4}, and consider a numerical
example for bound states of a ``small'' black hole of mass $10^{-11}$ times the
Earth mass (third part of our investigation).  This parameter combination leads
to gravitational electronic bound states 
(the coupling constant $\alpha_{G,{\rm BH}}$ given in 
Eq.~\eqref{alphaGBH} is small against unity).
It is thus possible to compare to a classical
treatment for circular Rydberg states, in terms of the trajectory shown in
Fig.~\ref{fig1}.  In the nonrelativistic approximation, the circular symmetry
(Schr\"{o}dinger approximation) is restored, while the relativistic
corrections, including the Fokker precession term (spin-orbit coupling term)
enter the relativistic energies given in Eqs.~\eqref{ea} and~\eqref{eb}.

In our investigations, we clarify, in particular, that the quantum 
mechanical gravitational central-field 
problem has quantum mechanical bound states.
This result holds  in the framework of
curved space-times (general relativity, see Ref.~\cite{Ei1916})
and takes into account the fact that it is impossible,
in contrast to the Dirac--Coulomb problem, to simply 
insert the gravitational potential $(-G m_1 m_2/r)$ into the 
Dirac Hamiltonian by the corresponding principle.
We evaluate the 
fine-structure formula for the Dirac--Schwarzschild Hamiltonian
[see Eq.~\eqref{mainres}],
and calculate the $\alpha_G^4$ corrections to the energy.
The bound-state energies are obtained as a 
function of ``good'' quantum numbers.

Let us briefly comment on the appropriate quantum 
numbers for the Dirac--Schwarzschild problem.
Because of the symmetries of the problem~\cite{Je2013,JeNo2013pra},
the principal quantum number $n$, 
the total angular momentum quantum number~$j$, 
and the Dirac angular momentum quantum 
number $\varkappa$ constitute a set of ``good'' quantum numbers.
The familiar spin-angular $ \chi_{\varkappa\,\mu}(\hat r) $ 
is assembled from the fundamental spinors and 
the spherical harmonics as follows~\cite{Ro1961,SwDr1991a,VaMoKh1988}.
It has the property
\begin{equation}
(\vec \sigma \cdot \vec L + 1) \, \chi_{\varkappa\,\mu}(\hat r) = 
-\varkappa \, \chi_{\varkappa\,\mu}(\hat r) \,.
\end{equation}
In the conventions of Refs.~\cite{Ro1961,SwDr1991a}, we have
$\varkappa = (-1)^{j+l+1/2} \, \left(j + \frac12 \right)$.
Knowing $j$ and $\varkappa$, one may calculate the 
orbital angular momentum quantum number $\ell = |\varkappa + 1/2|-1/2$
even if the orbital angular 
momentum operator $\vec L$ itself does not commute with the 
Dirac--Schwarzschild Hamiltonian~\eqref{HDS}.
Because $\varkappa$ can be mapped onto the 
orbital angular momentum quantum number $\ell$ (i.e.,
onto the ``spin orientation 
with respect to the orbital angular momentum''),
the main result~\eqref{mainres} is consistent.

For both the scalar Dirac Hamiltonian~\eqref{HSP} 
as well as the Dirac--Coulomb Hamiltonian~\eqref{HDC},
the explicit $\ell$ dependence of the spin-orbit coupling 
accidentally cancels out against the 
``implicit'' $\ell$ dependence of the 
matrix elements of the momentum, and the position 
operator [see Ref.~\cite{BeSa1957} and Eq.~\eqref{fs}].

%
%
\section*{Acknowledgments}

The research has been supported by the National Science Foundation 
(Grant PHY--1068547).

\appendix 

%
%
\section{Dirac--Coulomb Hamiltonian}
\label{appa}

For comparison, we 
briefly recall the Dirac--Coulomb Hamiltonian~\cite{BjDr1964,SwDr1991a}
\begin{equation}
H = \vec \alpha \cdot \vec p + 
\beta \, m_e - \frac{Z\alpha_{\rm QED}}{r} \,,
\end{equation}
where $Z$ is the nuclear charge number, and
$\alpha_{\rm QED} \approx 1/137.036$ is the QED fine-structure 
constant. The nonrecoil approximation is employed.
After a Foldy--Wouthuysen transformation,
the Hamiltonian takes the form
\begin{align}
\label{HDC}
H_{\rm DC}=& \;
\frac{\vec p^{\,2}}{2m_e} 
- \frac{Z\alpha_{\rm QED}}{r} 
\\[0.133ex]
& \; - \frac{\vec p^{\,4}}{8 m_e^3} 
+ \frac{\pi \, Z\alpha_{\rm QED}}{2 m_e^2} \, \delta^{(3)}(\vec r) 
+ \frac{Z\alpha}{4 m_e^2 r^3} \, \vec \Sigma \cdot \vec L \,.
\nonumber
\end{align}
The scaling corresponding to Eqs.~\eqref{rrho} and~\eqref{rrho_scalar}
reads as follows,
\begin{align}
r =& \; \frac{\hbar}{m_e c} \, \rho \,,
\quad
\vec\nabla_r = \frac{m_e c}{\hbar} \, \vec\nabla_\rho \,,
\quad
\vec p = -\ii \, \frac{m_e c}{\hbar} \, \vec\nabla_\rho \,.
\end{align}
The familiar~\cite{SwDr1991a,SwDr1991b,SwDr1991c}
scaled Dirac--Coulomb Hamiltonian is obtained as
\begin{align}
& H_{\rm DS} = \alpha_{\rm QED}^2 m_e  c^2 \, \left( -\frac12 \, \vec\nabla^2_\rho -
\frac{1}{\rho} \right)
\nonumber\\[0.133ex]
& \; + \alpha_{\rm QED}^4 \, m_e \, c^2 \, \left(
- \frac18 \, \vec\nabla^4_\rho
+ \frac{\pi}{2} \, \delta^{(3)}(\vec \rho)
+ \frac{\vec\sigma \cdot \vec L}{4 \, \rho^3} \right) \,.
\end{align}
The energy levels are given as follows~\cite{BjDr1964,SwDr1991a} 
\begin{align}
\label{fs}
E_{n \ell j} =& \;
-\frac{\alpha_{\rm QED}^2 \, m_e c^2}{2 n^2} 
\nonumber\\[0.133ex]
& \; + \alpha^4_{\rm QED} \, m_e c^2 \,
\left( \frac{3}{8 n^4} - \frac{1}{n^3 \, (2 j + 1)} \right) \,.
\end{align}
When evaluating the matrix elements according to 
formulas given on
pp.~15--17 of Ref.~\cite{BeSa1957}, one first obtains a 
functionally different formulas for $j=\ell + 1/2$ as opposed
to $j = \ell - 1/2$ but they coincide for given $j$.
This is analogous to Eq.~\eqref{mainres_scalar}.

\end{document}